\documentclass[useAMS,usenatbib]{mn2e}
\usepackage{times}

\input epsf
\usepackage{graphicx}
\newcommand{\beq}{\begin{equation}}
\newcommand{\beqa}{\begin{eqnarray}}
		  \newcommand{\eeq}{\end{equation}}
\newcommand{\eeqa}{\end{eqnarray}}

\newcommand{\vect}[1]{\mbox{\boldmath${#1}$}}
\newcommand{\lmk}{\left(}
\newcommand{\rmk}{\right)}

\newcommand{\lkk}{\left[} 
\newcommand{\rkk}{\right]}

\newcommand{\ven}{\vect n}

\newcommand{\vel}{{\vect l}}

\newcommand{\eps}{\epsilon}

\title[PDF for inclinations of merging compact binaries]{ Probability distribution function for inclinations of merging compact binaries detected by 
 gravitational wave interferometers} 
\author[N. Seto ]{Naoki Seto
\\
Department of Physics, Kyoto University, 
Kyoto 606-8502, Japan
}

\begin{document}

\maketitle
%
%
%
\begin{abstract}
We analytically discuss  probability distribution 
function (PDF) for inclinations of merging compact binaries whose gravitational waves 
are coherently  detected by 
a network of ground based interferometers.  The PDF would be  useful for 
studying  prospects of (1) simultaneously detecting electromagnetic signals (such 
as gamma-ray-bursts) associated with binary mergers and (2) statistically
constraining the related theoretical models from the actual observational data 
of multi-messenger astronomy.
Our approach is similar to Schutz (2011), but we explicitly include the dependence of 
the polarization angles of the binaries, based on the concise formulation 
given in  Cutler and Flanagan (1994). 
We find that the overall profiles of the 
PDFs are similar for any networks composed by the second 
generation detectors (Advanced-LIGO, Advanced-Virgo, KAGRA, LIGO-India).
For example, $5.1\%$ of detected binaries would have inclination angle less 
than $10^\circ$ with at most $0.1\%$ differences between the potential networks.
 A perturbative expression is also provided for generating the PDFs with a small 
number of 
parameters given by  directional averages of the quantity $\eps$ that 
characterises the asymmetry of network sensitivities to incoming
two orthogonal polarization modes.

\end{abstract}

\begin{keywords}
gravitational waves---binaries: close 
\end{keywords}

\section{Introduction}

Gravitational waves (GWs) from merging neutron star binaries 
(NS-NSs) are the most promising targets of ground-based detectors. For the 
upcoming second generation interferometers, the estimated detection rate of NS-NSs is 1-100/yr, and it is likely 
that we can succeed to directly detect their GWs within five years (Abadie 2010).
This estimated rate for NS-NSs is an order of magnitude higher than that for  black 
hole-neutron star binaries (BH-NSs), which also have relevance to this paper.

Meanwhile,  merging NS-NSs (and BH-NSs) are  strong candidates for  progenitors
of short gamma ray bursts (SGRBs) (see {\it e.g.} Nakar 2007; Berger 2013). Reflecting   geometry of the precedent inspiral phase,  
a merger product would have nearly axisymmetric profile around the direction 
$\vel$ of the orbital angular momentum of the binary (Metzger 
\& Berger 2012). If the progenitor scenario 
 for SGRBs
is the case, jet like structures would be launched soon after the merger, toward
the  polar directions $\pm \vel$, and they would be responsible for the observed gamma 
ray emissions. Later, more isotropic electromagnetic (EM) radiation might be 
emitted at lower energy band  as recently 
discovered for GRB130603B (Tanvir et al. 2013; Berger, Fong
\& Chornock 2013; Hotokezaka et al. 2013).

Therefore, searches for EM signals triggered by GW detections of compact binary inspirals would become an exciting field of astronomy, and various possibilities have been 
 actively discussed these days (see {\it e.g.} Fairhurst 2011; Schutz 2011; 
 Cannon et al. 2012; Evans et al. 2012; LIGO Scientific Collaboration 2013;  Nissanke, Kasliwal 
\& Georgieva 2013; Dietz et al. 2013; Kelley, Mandel  
\& Ramirez-Ruiz 2013; Piran, Nakar
\& Rosswog 2013; Ghosh \& Bose 2013; Kyutoku, Ioka 
\& Shibata 2014; Arun et al. 2014; Kyutoku \& Seto 2014). Given the expected axisymmetric profile of the merger
 products, it would be meaningful to evaluate the probability distribution 
 function (PDF) of inclinations for compact binaries whose GWs are detected by 
 the  second 
 generation detectors. Using the expected PDF, we can make statistical arguments about the future 
 prospects for simultaneously detecting EM signals and constraining theoretical models 
 based on observational data.

For a network of GW interferometers, the SNR of a binary depends on its sky 
direction $\ven$ and orientation $\vel$ (Cutler \& Flanagan 1994; Sathyaprakash 
\& Schutz 2009). 
Numerical studies by Monte Carlo simulations have been performed to
 properly
deal with these multi-dimensional angular parameters  
 (see {\it e.g.} Nissanke et 
al. 2010, 2013).

In this paper, we analytically study the PDFs of inclinations. Our underlying approach   is similar to Seto (2014) in which the 
relative detection rates of merging binaries were formally examined for general 
networks of detectors, but with no attention to  the PDFs of inclinations.
 Schutz (2011) discussed these two issues together,  by introducing certain approximation to 
the dependence of the polarization angles $\psi$ (explained in the next section) 
of binaries. But, for the PDFs of inclinations, the  accuracy of this 
approximation has not been clarified so far.   With the help of a concise 
expression provided by Cutler and Flanagan (1994),  our analysis does not rely on the approximation and thus can be 
used to study validity of the convenient method by Schutz (2011), as 
demonstrated below. 
Here the key quantity is $\eps(\ven)$ which
characterizes relative sensitivities of a network to two orthogonal polarization modes of incoming GWs.

Our analytical expressions derived in this paper are easily applicable to any
networks of ground-based interferometers.  We show that, in general, the PDFs 
depend weakly on  networks, especially for nearly face-on binaries. 
{ This is because the emitted GW power is strongest to the face-on direction 
for which the quantity $\eps(\ven)$ becomes less important, since the amplitudes of 
the two orthogonal polarization modes are nearly the same.  

In  contrast, for edge-on 
binaries, the PDFs depend strongly  
on $\eps(\ven)$ and have largest scatters, when comparing different networks.  However, the emitted GW power (and the 
detectable volume) is smallest for the
edge-on binaries.  Therefore,  among the sample of the detected  merging 
binaries, the relative fraction of the edge-on binaries  is much smaller than the face-on 
binaries.}

This paper is organized as follows.  In Sec.2, we explain our basic formulation, 
 assuming a coherent signal analysis for GWs from compact binary inspirals. 
We relate the total SNR and the expected detection 
rate of binaries. 
In Sec.3, we 
evaluate the PDF for inclinations of binaries at a given sky direction. 
In 
Sec.4, we discuss the full PDFs, including the angular averages with respect to 
sky directions. We also evaluate the PDFs concretely for  the planned 
second-generation interferometers. Then we  mention  relative detection rates 
of merging binaries, in relation to Seto (2014).
Sec.5 is devoted to a brief summary of this paper.

\section{Formulation}
\subsection{Signal-to-Noise Ratio}
Let us consider a binary at a sky direction $\ven$.  We use the unit vector $\vel$ for  the orientation of 
its orbital angular momentum.  This orientation vector is geometrically 
characterized by the two 
parameters $\theta$ and $\psi$. Here the inclination angle $\theta$ is the  angle between 
$\ven$ and $\vel$,  and the 
polarization angle $\psi$ fixes the rotational degree of freedom of $\vel$ 
around the line-of-sight $\ven$ (Cutler \& Flanagan 1994; Sathyaprakash 
\& Schutz 2009).

In the principle polarization frame of the binary, the two polarization modes 
$+$ and $\times$ of the mass-quadrupole waveform are proportional to $d_+$ and 
$d_\times$ given by 
\beq
d_+(I)=\frac{I^2+1}2,~~d_\times(I)=I \label{amp1}
\eeq
with $I\equiv \cos \theta$ ($I=1$ for face-on and $I=0$ for edge-on; Peters \& 
Mathews 1963). Below, we simply term $I$  inclination.
In astrophysical context, we are not interested in the sign of $I$ and 
hereafter consider its absolute value (namely $0\le I\le 1$). Correspondingly, 
the inclination angle  $\theta$ is limited to the range $0\le \theta \le 
90^\circ$ (identifying $\pi-\theta \to \theta$). 
Throughout this paper, we neglect the  precessions of orbital planes of 
binaries due to their spins. This would be a reasonable approximation for NS-NSs 
whose orbital angular momenta would  dominate the spin angular momenta, 
 due to their comparable masses and expected spin parameters much smaller than 
 those of  black holes 
(Cutler \& Flanagan 1994; Apostolatos et 
al. 1994).

For  detecting GWs from binaries, we consider to make coherent signal analysis 
using totally $m$ ground-based interferometers  with no correlated detector noises.   Reflecting the spin-2 nature of 
GWs, the responses of each interferometer (labeled with $i$) to the 
two polarization  modes are written by
\beqa
c_{i+}(\ven,\psi)&=&a_i(\ven)\cos2\psi+b_i(\ven)\sin2\psi,\label{cp}\\
c_{i\times}(\ven,\psi)&=&-a_i(\ven)\sin2\psi+b_i(\ven)\cos2\psi.\label{cc}
\eeqa
The explicit forms of the functions $a_i(\ven)$ and $b_i(\ven)$ can be found in 
Schutz (2011) 
(see  Eqs.(19) and (20) therein). Here the overall amplitude of $(a_i,b_i)$ is 
proportional to the so-called horizon distance of the detector $i$.\footnote{The 
horizon distance is the detectable range of a gravitational wave source that is 
optimally located and oriented. For a NS-NS binary of $1.4M_\odot+1.4M_\odot$ 
with the detection threshold of $SNR=8$, each advanced-LIGO interferometer 
is planned to have the horizon distance of 445Mpc (Abadie et al. 2010). Advance-Virgo and KAGRA would 
have similar values. 
} 
But, below, 
we simply assume that all the interferometers have an identical noise curve with 
 the same horizon distance.  
 In practice, it is straightforward to take into account the differences of the horizon 
 distances  by setting appropriate weights for
 the functions $(a_i,b_i)$.

For the coherent signal analysis, the total signal-to-noise ratio (SNR) is obtained 
from Eqs.(\ref{amp1})-(\ref{cc}) and 
depends on the geometrical parameters $(\ven,I,\psi)$ as 
\beqa
SNR^2\propto \sum_{i=1}^m\lkk \lmk c_{i+}d_+    \rmk^2+\lmk c_{i\times}d_\times    
\rmk^2   \rkk \equiv f(\ven,I,\psi)\label{snr}
\eeqa
(Cutler \& Flanagan 
1994; Dietz et al. 2013).
Note that we only included the lowest quadrupole mode (\ref{amp1}) for 
estimating the  total SNR. This would 
be a good approximation for NS-NSs, since the next order correction is 
proportional to the mass difference and NS-NSs are expected to have similar 
masses, as commented earlier (Van Den Broeck 
\& Sengupta 2007; Blanchet et 
al. 2008; Tagoshi et al. 2014).  In Eq.(\ref{snr}), SNR is inversely proportional to the distance to 
the binary, while we omitted its explicit dependence.

\begin{figure}
\begin{center}
\includegraphics[width=10.cm,clip]{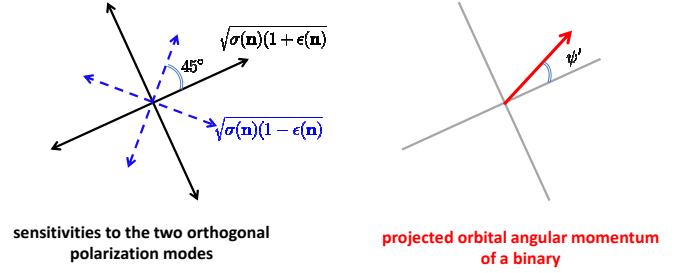}
\end{center}

\vspace*{-0.2cm}

\caption{ The geometric interpretation of Eq.(5) for incoming GW from a sky direction 
$\ven$. 
(Left panel) In the plane normal to $\ven$, the network has  two  
orthogonal polarization bases at specific orientations, and measure these two 
modes with  sensitivities   proportional to $\sqrt{\sigma(\ven)(1+\eps(\ven))}$ 
and  
$\sqrt{\sigma(\ven)(1-\eps(\ven))}$. Here the parameter $\sigma(\ven)$ 
represents the total sensitivity to the two modes and $\eps(\ven)$ shows the 
asymmetry between them.  
(Right panel) The orbital angular momentum of the binary is  projected 
to the normal plane. Its orientation is  characterised by the angle $\psi'$  measured from the better 
sensitivity mode in the left panel. The original amplitudes (1) are given for 
the polarization modes symmetric to this projected vector.
}
\label{t3}
\end{figure}

With trigonometric relations (Cutler \& Flanagan 1994), the $\psi$-dependence of $f$ is simplified as
\beq
 f(\ven,\psi,I)=   \sigma(\ven) \lkk D_0 
 +\epsilon(\ven) D_1\cos4\psi'   \rkk \label{deff},
\eeq
where the new polarization angle $\psi'=\psi-\delta(\ven)$ is related to the original ones $\psi$
with an offset $\delta(\ven)$ that satisfies the following relation
\beq
\tan \delta (\ven) = \frac{2\sum_{i=1}^m a_i(\ven)b_i(\ven)}{\sum_{i=1}^m (a_i(\ven)^2-b_i(\ven)^2)}.
\eeq

In Eq.(\ref{deff}), the functions $D_0$ and $D_1$ are defined by
\beq
D_0(I)\equiv  (d_+^2+d_\times^2)=\frac{I^4+6I^2+1}{4} 
\eeq
\beq
D_1(I)\equiv  (d_+^2-d_\times^2)=\frac{(I^2-1)^2}{4} 
\eeq
and the parameters $\sigma$ and $\eps$ are defined by
\beqa
\sigma(\ven) &\equiv&  \sum_{i=1}^m 
a_i^2+b_i^2\\
\epsilon(\ven)&\equiv &\frac{\sqrt{\lkk \sum_{i=1}^m (a_i^2-b_i^2)\rkk^2+4(\sum_{i=1}^m a_ib_i)^2}}{\sigma(\ven)}.
\eeqa

From Cauchy-Schwartz inequality, we have 
\beq
0\le \epsilon(\ven)\le 1.
\eeq
 The 
equality $\eps(\ven)=1$ holds only when the vector $(a_1(\ven),\cdots,a_m(\ven))$ is 
parallel to $(b_1(\ven),\cdots,b_m(\ven))$, including the case for a single detector network 
 (with identity $\eps(\ven)=1$ for all the directions $\ven$).

The geometric meaning of Eq.(5) is explained in Fig.1. 
The parameter $\eps(\ven)$ characterizes the asymmetry of 
network sensitivity to the two  orthogonal polarization modes given for each 
direction $\ven$ (Cutler \& Flanagan 1994). This parameter plays an important 
role in this paper.
With respect to the polarization decomposition shown in the left panel, 
  the amplitudes of the quadrupole 
waves of the binary (with the projected angle $\psi'$) are given by  
\beq
\lmk d_+^2 \cos^2 
2\psi'+d_\times^2 \sin^2 2\psi' \rmk^{1/2}
\eeq
 and  
\beq
\lmk d_+^2 \sin^2 
2\psi'+d_\times^2 \cos^2 2\psi' \rmk^{1/2}
\eeq
 with $d_+$ and $d_\times$ 
defined by Eq.(1).
Then we have
\beqa
SNR^2&\propto& \sigma(\ven)(1+\eps(\ven))\lmk d_+^2 \cos^2 
2\psi'+d_\times^2 \sin^2 2\psi' \rmk\nonumber \\
 & &+\sigma(\ven)(1-\eps(\ven))\lmk d_+^2 \sin^2 
2\psi'+d_\times^2 \cos^2 2\psi' \rmk \nonumber.
\eeqa
The right-hand-side of this relation is identical to that of  Eq.(5).

For a face-on binary $I=1$, we have $d_+=d_\times$ (thus $D_1=0$) and this expression does not 
depend on the angle $\psi'$ (and $\eps$).
On the other hand, edge-on binaries 
($I=0$)
emit $100\%$ linearly polarized GWs and Eq.(5) depends strongly on $\eps(\ven)$ 
and $\psi'$ with $D_0=D_1=1/4$.

\subsection{Detectable Binaries}

We define  $r_{max}$ as the maximum distances to the binaries detectable  above a given 
SNR threshold.  Then, from Eq.(\ref{snr}), we have a scaling relation
\beq
r_{max}\propto f(\ven,I,\psi)^{1/2}
\eeq
(Finn \& Chernoff 1993; Schutz 2011; Dietz et al. 2013).
Therefore, assuming that merging binaries have random orientations and spatial 
distributions, the expected number of detectable ones in a parameter range 
$d\ven dId\psi$ is proportional to
 \beq
 f(\ven,I,\psi)^{3/2} d\ven d\psi dI .
\eeq
Here we neglected cosmological effects that would be unimportant  
at least for NS-NSs observed with second generation detectors.
In this paper, we  study the PDFs  in appropriately 
 normalized forms. Therefore the actual values of the horizon distance and the comoving 
 merger rate are irrelevant to our results.

Next we integrate out the less interesting polarization parameter $\psi$ and define 
the new function $\alpha(\ven,I)$ by
\beq
\alpha(\ven,I)\equiv\frac2\pi \int_0^{\pi/2} f(\ven,I,\psi)^{3/2} d\psi.\label{al}
\eeq

As we initially integrate  the polarization angle $\psi$ (or equivalently $\psi'$) before 
integrating the sky direction $\ven$, 
we actually do not need to directly handle the complicated offset $\delta 
(\ven)$. This is an advantageous point of our approach, and simplifies the 
actual evaluation of PDFs.

From Eq.(\ref{deff}), the integral $\alpha(\ven,I)$ can be formally expressed as
\beq
\alpha(\ven,I)=\sigma(\ven)^{3/2} D_0(I)^{3/2} \gamma \lkk \eps(\ven)R(I)\rkk \label{defal},
\eeq
where we define
\beq
R(I)\equiv \lmk\frac{D_1}{D_0}  \rmk=\frac{(I^2-1)^2}{I^4+6I^2+1}\label{ri}
\eeq
and
\beq
\gamma(x)\equiv \frac2\pi \int_0^{\pi/2} \lmk 1+x \cos4\psi \rmk^{3/2}d\psi.
\eeq

The integral $\gamma(x)$ is given as follows
\beq
\gamma(x)=\frac{2(1-x)^{1/2}}{3\pi} \lkk 4 E\lmk \frac{2x}{x-1}   \rmk-(1+x) K\lmk \frac{2x}{x-1}   \rmk  \rkk 
\eeq
with the incomplete elliptic integral of the 
second kind $E(x)$ and  the complete elliptic integral of the first kind $K(x)$ 
defined respectively  by  
\beq
E(x)\equiv\int_0^{\pi/2}(1-x \sin^2\theta)^{1/2}d\theta
\eeq
\beq
K(x)\equiv \int_0^{\pi/2}(1-x \sin^2\theta)^{-1/2}d\theta
\eeq
(see also Dietz et al. 2013).

Around $x=0$,  the integral $\gamma(x)$ is expanded as follows;
\beqa
\gamma(x)&=&1+\frac{3}{16} x^2+\frac{9}{1024} x^4+\frac{35}{16384} x^6+\frac{3465}{4194304} x^8\nonumber\\
& &+\frac{27027}{67108864} x^{10}+\frac{969969}{4294967296} x^{12}+O(x^{14}).\label{gep}
\eeqa
We use this expression later in Sec.4.

Finally, after integrating the sky direction $\ven$ of binaries, the PDF for a network 
can be formally expressed as
\beq
P_{net}(I)=\frac{\displaystyle \int_{4\pi}d\ven \,\alpha(\ven,I)}{\displaystyle\int_{0}^1dI\int_{4\pi}d\ven\, \alpha(\ven,I)}.\label{pnet}
\eeq
Here the denominator is a normalization factor to realize
\beq
\int_{0}^1P_{net}(I)dI=1.
\eeq

\subsection{Total Detection Rate}

Our formulation up to Eq.(15) is similar to Seto (2014) in which the relative 
detection rates of binaries were examined by integrating all the angular 
variables including  $I$, without paying 
attention to its PDF.

In this subsection,  we define the following quantity
\beq
X\equiv \int_0^1dI\int_{4\pi}d\ven \alpha(\ven,I), \label{intr}
\eeq
and briefly summarize the arguments in Seto (2014).
Here we only extracted geometrical information relevant for the relative 
detection rates, considering comparisons
 between different networks.
 Actually, the integral (\ref{intr}) for the relative rates is 
identical to the denominator in Eq.(24).

For a hypothetical network with $\epsilon(\ven)=0$, the function 
$\alpha(\ven,I)$ becomes a separable form as $\sigma(\ven)^{3/2}D_0(I)^{3/2}$ 
and we have
\beq
X_0\equiv \int_0^1dI D_0(I)^{3/2}\int_{4\pi}d\ven 
\sigma(\ven)^{3/2}=N_0\int_{4\pi}d\ven \sigma(\ven)^{3/2}
\eeq
with the parameter $N_0\equiv 0.82155$.
This expression can be easily evaluated and we do not need to directly deal with the
dependence on the orientation angles $(I,\psi)$ of binaries. Therefore, as a 
convenient approximation to the original complicated one $X$, we might use $X_0$ for general networks with 
$\epsilon(\ven)\ne 0$. Indeed, the expression $X_0$ is essentially the same as 
that proposed by Schutz (2011) for estimating the relative rates.

The question here is how 
well the original integral $X$ is reproduced by the approximation $X_0$.
In order to check this,  we define the ratio
\beq
Y\equiv \frac{X}{X_0}.
\eeq
The main result in Seto (2014) is the following relation 
\beq
Y=\frac{\int_{4\pi}d\ven\ \sigma(\ven)^{3/2}G[\epsilon(\ven)]}{\int_{4\pi}d\ven \ 
\sigma(\ven)^{3/2}}
\eeq
where $G(x)$ is a monotonically increasing function of $x$ and perturbatively 
expanded as
\beq
G(x)=1+0.00978x^2+0.00026x^4+O(x^6)
\eeq
 with $G(0)=1$ and 
$G(1)=C_{01}\equiv 1.010125$. 

Given the inequality $\sigma(\ven)\ge 0$, we generally have the bounds
\beq
1\le Y\le 1.010125.\label{gb}
\eeq
Therefore the simple expression $X_0$ is an excellent approximation to $X$.
These inequalities would be practically sufficient for astronomical arguments, 
but,  we can actually evaluate the ratio $Y$, as a byproduct of our perturbative 
formulation. This will be discussed in Sec.4.5.

\begin{figure}
\begin{center}
\includegraphics[width=8.cm,clip]{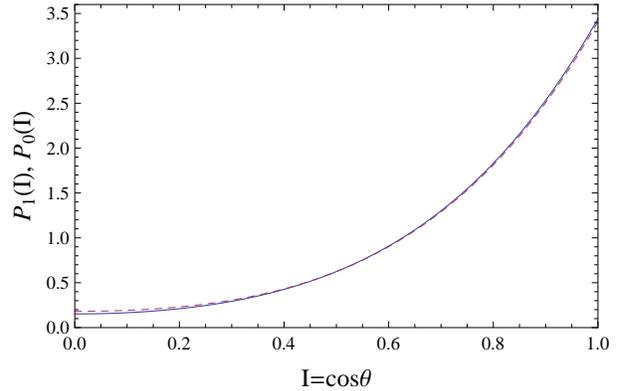}
\end{center}

\vspace*{-0.2cm}

\caption{The functions $P(I,\eps)$ for $\eps=0$ 
and 1.  The solid curve represents 
$P_0(I)=P(I,0)$ and the dashed one is for $P_1(I)=P(I,1)$. The latter is 
identical to
the sky averaged PDF for a single detector. 
}
\label{t3}
\end{figure}

\section{PDFs for given sky directions}
In this section, we discuss the PDFs of inclinations $I$ for a fixed parameter 
$\eps$,  
without taking the sky average as in Eq.(\ref{pnet}). From Eq.(\ref{defal}),  we define the function $P(I,\eps)$ 
as follows
\beq
P(I,\eps)=\frac{D_0(I)^{3/2}\gamma[\eps R(I)]}{N_\eps}
\eeq
with the normalization factor
\beq
N_\eps=\int_{0}^1 dI D_0(I)^{3/2}\gamma[\eps R(I)].
\eeq
The function $P(I,\eps)$  $(0\le \eps \le 1)$  can be regarded as the PDF for a 
given sky direction $\ven$ with $\eps(\ven)=\eps$.  In addition, for the special 
value $\eps=1$, it corresponds to the full (sky averaged) PDF for a
network composed by a single interferometer that identically has $\eps(\ven)=1$, 
 as mentioned earlier. 
Our primary task in this section is to explicitly demonstrate that the function 
$P(I,\eps)$ does not have
 strong dependence on $\eps$.

To begin with, we introduce the notations $P_0(I)$ and $P_1(I)$ for the two boundary 
parameters $\eps=0$ and 1 by 
\beq
P_0(I)\equiv P(I,0)= \frac{D_0(I)^{3/2}}{N_0}\label{p0}
\eeq
\beq
P_{1}(I)\equiv P(I,1)= \frac{D_0(I)^{3/2}}{N_{1}} \gamma\lkk R(I) \rkk \label{p1}
\eeq
with the normalization factors
$N_0=0.82155$ (already appeared in Eq.(27)) and $N_1=0.82986$. We have $ 
N_1/N_0=C_{01}=1.010125$. 

Schutz (2011) studied the PDF of inclinations for detected binaries. He used an 
approximation in which the explicit $\psi$ dependence
was not included for the effective volume (\ref{deff}). In our language, this treatment corresponds to commute the 
order of the following two operations in Eq.(\ref{al});  (i) the nonlinear 
manipulation $[\cdots]^{3/2}$ and (ii) the $\psi$-averaging. It is equivalent to taking $\eps(\ven)=0$ in Eq.(\ref{defal}).  Consequently, his PDF is 
identical to $P_0(I)$ defined in Eq.(\ref{p0}). 
In this paper, we can analytically show that this PDF 
generally  serves as a  good approximation, irrespective of the details of a network.

In Fig.2 we present $P_0(I)$ (solid curve) and $P_1(I)$ (dashed curve).  The two 
curves show similar shapes.   In order to 
enhance the differences between them,  we show the ratio 
$P_1(I)/P_0(I)$ (dashed curve) in Fig.3,  together with $P(I,\eps)/P_0(I)$ at the 
intermediate values $\eps=0.1,0.2,\cdots,0.9$ (solid curves).

For a given $\eps$,   the function $P(I,\eps)$ becomes minimum at $I=0$, 
reflecting the smallest amplitude at the edge-on configuration. 
At the same time, as shown in Fig.3,  the ratios  $P(I,\eps)/P_0(I)$ show the largest scatter 
at $I=0$.  This is because the emitted waves are 100\% linearly polarized and 
the effects of the asymmetry parameter $\eps$ become significant.

In contrast, at 
the face-on configuration $I=1$, we obtain
$R(I)=0$ and $\gamma (\eps R(I))=1$.  Therefore, around $I\sim 1$, we approximately have
\beq
P(I,\eps)\simeq \frac{D_0(I)^{3/2}}{N_\eps}
\eeq
with $P(I,\eps)/P_0(I)\simeq N_0/N_\eps$ that is now a decreasing function of 
$\eps$ with the minimum value $N_0/N_1=1/C_{01}=0.99$  at $\eps=1$. This shows 
that, around $I\sim 1$, the relative difference between $P(I,\eps)$ is at most $\sim 1\%$.

As shown in Fig.3, the two functions $P_0(I)$ and $P_1(I)$ intersect at 
$I=0.489$ where  the family
$P(I,\eps)$ depends very weakly on $\eps$.
Except the tiny region around this intersection, the function $P(I,\eps)$ 
($0\le \eps \le 1$) is bounded by the two curves $P_0(I)$ and $P_1(I)$. 
In the next section, we apply this result for discussing the overall profile of the 
sky-averaged function $P_{net}(I)$.

So far, we have studied the PDFs  for  $I$ only in a differential form.  Here we examine 
the cumulative PDFs ${\cal P}_{cum}(\theta,\eps)$ for the inclination angle $\theta=\cos^{-1 }I$ 
defined  by
\beq
{\cal P}_{cum}(\theta, \eps)\equiv \int^{1}_{\cos \theta} P(I,\eps)dI\label{cpc}
\eeq
with $0\le \theta\le 90^\circ$.
This function represents the probability that a detected binary has a viewing angle less 
than $\theta$, from its symmetry axis $\vel$.
 In this cumulative form, we rigidly have the following bounds 
\beq
 {\cal P}_{cum}(\theta,1)\le {\cal P}_{cum}(\theta,\eps) \le {\cal 
 P}_{cum}(\theta,0) \label{tight}
\eeq
and the two boundaries have small relative differences 
\beq
1\le {\cal P}_{cum}(\theta,0)/{\cal P}_{cum}(\theta,1)\le C_{01}=1.010125.
\eeq
We can confirm their similarity in Fig.4. The tight confinement (\ref{tight}) 
would become useful  in the next section.

For conveniences at astronomical studies, we provide a fitting function for ${\cal P}_{cum}(\theta,\eps)$
\beqa
{\cal P}_{cum,f}(\theta)&=&4.23888\lmk  \frac{\theta}{90^\circ}  \rmk^2 
-0.373208\lmk  \frac{\theta}{90^\circ}  \rmk^3\nonumber \\
& &-6.64160 \lmk \frac{\theta}{90^\circ}  \rmk^4\label{pcumf}
\eeqa
which reproduces the functions ${\cal P}_{cum}(\theta,\eps)$ ($0\le \eps \le 1$) 
with relative error less than 1\% in the range
$0\le \theta \le 30^\circ$.
In Table.1, we also evaluate the mean $[{\cal P}_{cum}(\theta,0)+{\cal 
P}_{cum}(\theta,1)]/2$ for some representative angles $\theta$.

\begin{figure}
\begin{center}
\includegraphics[width=8cm,clip]{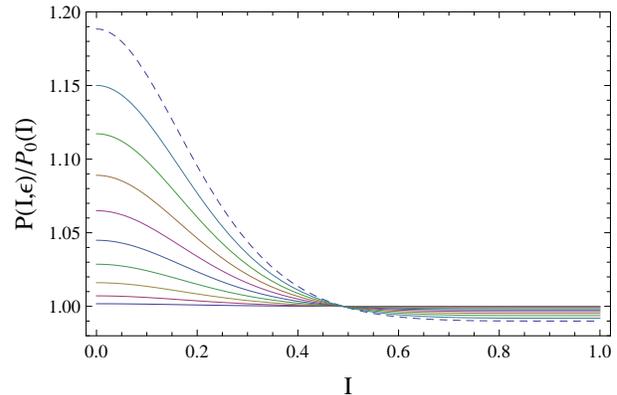}
\end{center}

\vspace*{-0.2cm}

\caption{The ratios $P_1(I)/P_0(I)$ (dashed curve) and  $P(I,\eps)/P_0(I)$ with 
$\eps=0.1,0.2,\cdots,0.9$ (solid curves from bottom to top). At $I=0$, the ratio $P(I,\eps)/P_0(I)$ is an
increasing function of $\eps$.  The two function $P_1(I)$ and $P_0(I)$ intersect  at $I=0.489$.
}
\label{t3}
\end{figure}

\begin{figure}
\begin{center}
\includegraphics[width=8.cm,clip]{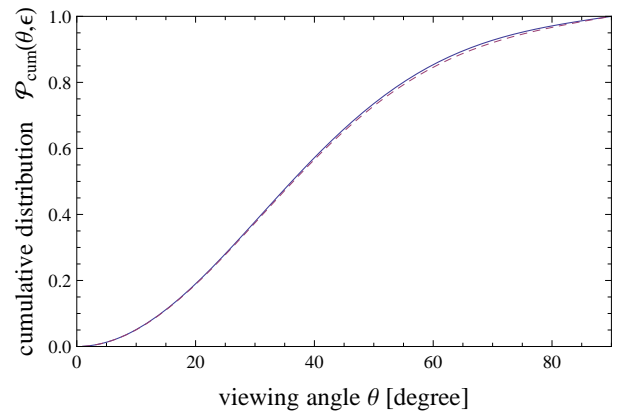}
\end{center}

\vspace*{-0.2cm}

\caption{The cumulative functions ${\cal P}_{cum}(\theta,\eps)$ for $\eps=0$ (solid 
curve) and 1 (dashed curve). 
Their relative difference is only $\sim 1\%$. For $0\le \eps\le 1$, the function ${\cal 
P}_{cum}(\theta,\eps)$ is tightly bounded by these two curves.
}
\label{t3}
\end{figure}

\begin{table}
\caption{The cumulative PDF: $[{\cal P}_{cum}(\theta,0)+{\cal 
P}_{cum}(\theta,1)]/2$ at sample points.}
  \begin{tabular}{|c|c|c|c|c|c|c|} \hline
    $\theta$ & $1^\circ$ & $5^\circ$ & $10^\circ$ & $20^\circ$ & $45^\circ$& $80^\circ$ \\ \hline 
 cumulative PDF &$ 5.2\times 10^{-4}$ & 0.013 & 0.051& 0.19 &0.66 &0.969 \\ \hline
  \end{tabular}
\end{table}

\section{all sky distribution}
In this section, we discuss the full (sky averaged) functions $P_{net}(I)$ defined in Eq.(\ref{pnet}) for 
various networks of ground-based interferometers. In Sec.4.1 we first mention their overall 
profiles based on the results shown in the previous section. Then, in Sec.4.2, we 
use the perturbative expansion (\ref{gep}) and derive an expression for more preciously 
evaluating $P_{net}(I)$. The validity of our perturbative method is examined in 
Sec.4.3.  In Sec.4.4,  we 
apply our method for networks composed by second generation interferometers. In 
Sec.4.5, we mention the relative detection 
rates of merging binaries, in relation to Seto (2014) and Sec.2.3.

\subsection{general remarks}

From Eqs.(\ref{defal}) and (\ref{pnet}), the function  $P_{net}(I)$ is obtained by taking an average of 
$P[I,\eps(\ven)]$ with the following relative weights
\beq
d\ven \sigma(\ven)^{3/2} N_{\eps(\ven)}\label{wei}.
\eeq
Therefore, similar to the previous one $P(I,\eps)$,  the averaged one $P_{net}(I)$ should be bounded by the two functions 
$P_0(I)$ and $P_1(I)$ except the tiny region around their intersection at 
$I=0.489$, as mentioned earlier in Fig.3. This means that the overall profile of 
$P_{net}(I)$ can be approximately understood from the shapes of the two functions $P_0(I)$ and 
 $P_1(I)$. Around $I\sim 1$, the function 
$P_{net}(I)$ weakly depends on the details of a network (see Fig.3). 
Among the binaries detected by a single interferometer, the fraction of nearly 
edge-on ones ($I\sim 0$) could be at most $\sim20\%$  larger than a network with multiple 
interferometers.

Next, we discuss the   cumulative PDFs for networks.  As in Eq.(\ref{cpc}), we 
define ${\cal P}_{cum,net}(\theta)$ by
\beq
{\cal P}_{cum,net}(\theta)\equiv \int^1_{\cos\theta}P_{net}(I)dI.\label{cnet}
\eeq
By changing the order of the integrals $d\ven$ and $dI$, we can understand that the function
${\cal P}_{cum,net}(\theta)$ is obtained by averaging 
$ {\cal P}_{cum}[\theta,\eps(\ven)]$ again with the weight (\ref{wei}). 
Since the cumulative PDFs $ {\cal P}_{cum}[\theta,\eps(\ven)]$ are tightly bounded by the two functions ${\cal 
P}_{cum}(\theta,0)$ and ${\cal P}_{cum}(\theta,1)$, the sky averaged one ${\cal P}_{cum,net}(\theta)$ must 
be also bounded by them. Therefore, with  relative error less 
than $\sim 1\%$, we can apply the previous fitting 
formula (\ref{pcumf})  for the sky averaged one ${\cal P}_{cum,net}(\theta)$ in the range $0\le 
\theta \le 30^\circ$, irrespective of the details of  networks. Similarly, we can 
apply Table.1 for given networks.

For example, $\sim 5\%$ of detected binaries have viewing angle $\theta$ less 
than $10^\circ$. The fraction becomes $\sim 1.3\%$ for $\theta \le 5^\circ$.
In other words, if one hundred binaries are detected by a network, the minimum 
inclination angle would be $\theta\sim 5^\circ$ and we will have $\sim 5$ 
binaries with $\theta$ less than $10^\circ$.

\subsection{perturbative evaluation}
Now we move to develop a perturbative method for  evaluating $P_{net}(I)$ more precisely.
First we rewrite $P_{net}(I)$ as follows
\beq
P_{net}(I)=\frac{Q_{net}(I)}{M_{net}}, \label{pnet2}
\eeq
where the numerator and the denominator are non-dimensional quantities defined by
\beq
Q_{net}(I)=\frac{\displaystyle \int_{4\pi}d\ven \,\alpha(\ven,I)}{\displaystyle\int_{4\pi}d\ven\, \sigma(\ven)^{3/2}},
~~M_{net}=\frac{\displaystyle \int_{0}^1dI\int_{4\pi}d\ven \,\alpha(\ven,I)}{\displaystyle\int_{4\pi}d\ven\, \sigma(\ven)^{3/2}}.\label{mnet0}
\eeq
Here we introduced the common factor $\lkk \int_{4\pi}d\ven\, 
\sigma(\ven)^{3/2}\rkk^{-1}$ to make our analysis comprehensive. 
Applying the expansion (\ref{gep}) for $Q_{net}(I)$, we obtain
\beq
Q_{net}(I)=D_0(I)^{3/2} \lmk  1+\frac{3 R^2 s_2}{16}  +\frac{9R^4 s_4}{1024} 
 +\frac{35R^6s_6 }{16384}+\cdots  \rmk\label{qnet}
\eeq
with a function $R(I)$ defined in Eq.(\ref{ri}) and  the coefficients $s_j$ 
given by
\beq
s_j\equiv \frac{\displaystyle \int_{4\pi} \sigma(\ven)^{3/2} \eps(\ven)^j d\ven}{\displaystyle\int_{4\pi} \sigma(\ven)^{3/2}  d\ven}.
\eeq
From the inequalities $0\le \eps(\ven) \le1$, we have
\beq
0\le s_{j+1}\le s_{j}\le1
\eeq
with the equality $s_j=s_{j+1}$ only for  $s_j=0$ (identically $\eps(\ven)=0$) or  
$s_j=1$ (identically $\eps(\ven)=1$).
In our perturbative approach, all the information of a network is projected into 
the sequence of numbers $(s_2,s_4,s_6,\cdots)$. We thus call them network parameters.

In the same manner, the normalization factor $M_{net}$ can be perturbatively evaluated as
\beq
M_{net}=N_0+\frac{3}{16} u_2 s_2 +\frac{9}{1024} u_4 s_4+\frac{35}{16384} u_6 s_6+\cdots\label{mnet}
\eeq
where we define the parameters $u_j$ given by the following integrals
\beq
u_j\equiv {\int_{0}^1 D_0(I)^{3/2} R(I)^j  dI} \label{uj}.
\eeq
These are constants and do not depend on  networks.
In Table.2, we present them up to $u_{12}$.

\begin{table}
\caption{The parameters defined in Eq.(\ref{uj}).}
  \begin{tabular}{|l|c|c|c|c|c|} \hline
    $u_2$ & $u_4$ & $u_6$ & $u_8$ & $u_{10}$ & $u_{12}$ \\ \hline 
 0.042885 & 0.024136 & 0.018303 & 0.015297& 0.013400 &0.012067 \\ \hline
  \end{tabular}
\end{table}

\subsection{expansion for a single detector network}

In the previous subsection, we explained how to perturbatively evaluate the sky 
averaged function $P_{net}(I)$.  Our expression (\ref{pnet2}) is characterized 
by the network parameters
$(s_2,s_4,\cdots)$ with $0\le s_j\le 1$.  From Eqs.(\ref{qnet}) and (\ref{mnet}), we will have better convergence for 
smaller $s_j$.  On the other hand, the convergence would become worst for the maximum 
value $s_j=1$, corresponding to a single detector network. But, for this case, 
we actually have the non-perturbative result $P_1(I)$ given in Eq.(\ref{p1}). 
Therefore,  we can test the validity of our perturbative expansion by comparing 
the two results.

\begin{figure}
\begin{center}
\includegraphics[width=8.cm,clip]{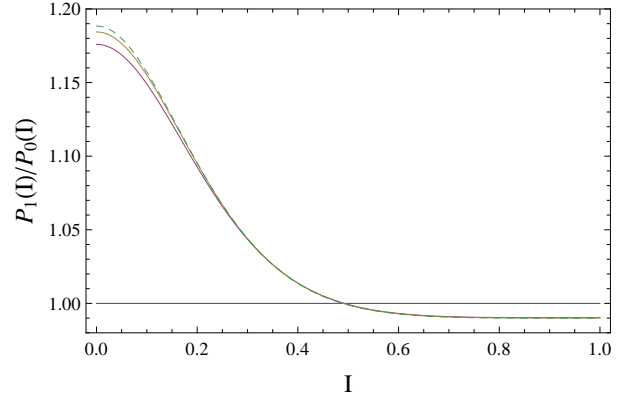}
\end{center}

\vspace*{-0.2cm}

\caption{The ratio $P_1(I)/P_0(I)$ (dashed curve) and its perturbative 
expansions according to Eq.(\ref{spt}) (0th, 2nd and 4th order approximations: 
solid curves from bottom). Convergence of the perturbative expansion is fast.
}
\label{t3}
\end{figure}

For $s_j=1$, our perturbative expression is given by
\beq
P_1(I)=D_0(I)^{3/2} \frac{   1+\frac{3}{16}  R^2  +\frac{9}{1024} 
R^4  +\frac{35}{16384} R^6+\cdots  }{N_0+\frac{3}{16} u_2  +\frac{9}{1024} u_4 
+\frac{35}{16384} u_6 +\cdots} \label{spt}.
\eeq
In Fig.5, we show the non-perturbative results (dashed curve) and the 0th, 2nd 
and 4th order approximations (solid curves). This figure shows that, even in 
the worst case $s_j=1$, the convergence is fast and the relative error is at most $\sim 
0.3\%$ with the 4th order approximation. Therefore, our perturbative method 
would be efficient to reproduce the function $P_{net}(I)$.

\subsection{second generation detector networks}
Now we concretely evaluate the averaged function $P_{net}(I)$ for networks of ground-based 
interferometers. We consider the following five second-generation interferometers; 
LIGO-Hanford (H), LIGO-Livingston (L), Virgo (V), KAGRA (K) and LIGO-India (I). For 
their locations and orientations, we use Table.2 in Schutz (2011).  But, for KAGRA, 
we apply the updated data; the geographical position ($137.3^\circ$E, $36.4^\circ$N) and the 
orientation angle $74.6^\circ$ for the bisector of its two arms measured counter-clock 
wise from the local East direction. All the detectors are assumed to have 
identical noise spectrum (and thus the identical horizon distance).

In Table.3, we present the network parameters $s_j$ for various potential networks 
composed by the five interferometers. We have the identities $s_j=1$ for single interferometer, 
as mentioned earlier.  The two LIGO interferometers H and L are separated by $\sim 
3000$km but configured to 
realize  large overlaps for incoming GW signals (Cutler \& Flanagan 1994). To this end, their orientations are nearly aligned. This results in larger network 
parameters $s_j$, compared with other two-detector networks such as HV or VK.

The five-detector network HLVKI has the smallest network parameters $s_j$ in  Table.3, indicating that 
due to the randomness of detector configurations, the degree of asymmetry $\eps$  decreases.

In Fig.5, we show the full functions $P_{net}(I)$ for a single interferometer 
(dashed curve) as well 
as the HL, HLV, HLVK and HLVKI networks (solid curves from top to bottom). We use the 12th order approximation for the 
perturbative expansion. Using {\it Mathematica}, we can straightforwardly 
calculate the network parameters $s_j$ and  evaluate the perturbative expressions. 
As we increase the number of interferometers, the PDF moves from $P_1(I)$ (for 
a single interferometer) to $P_0(I)$, decreasing  fraction of edge-on binaries.

{ The PDF for the HL network is close to that of a single interferometer, as 
easily expected from the relatively large network parameters in Table.3. For nearly edge-on binaries ($I\sim 0$), the 
number of detectable volume depends on $\psi'$ as $\propto 
[1+\eps(\ven)\cos4\psi']^{3/2}$} (see Eq.(5)), and the detected binaries are 
likely to have
  polarization angles  around  $\psi'=0$ (mod $\pi/2$) for 
the HL network. 
The PDFs for the HLVKI network is reproduced by  Schutz's approximation
$P_0(I)$ with error less than $8\%$, even around $I\sim 0$. 

The fraction of 
nearly edge-on binaries detected by  the HLVKI network would be $\sim 10\%$ 
smaller than that of the HL network. But we should recall that the emitted  GW 
power (thus the detectable range) is smallest to the edge-on direction $I\sim 0$. Indeed, 
we have the ratio of the emitted powers $D_0(I=0)/D_0(I=1)=1/8$, compared with  face-on binaries $I=1$. 
As 
shown in Fig.2, the nearly edge-on binaries would be a minor component in the 
whole detected sample. 

\begin{table}
\caption{The network parameters $s_j$ for various networks of ground-based 
interferometers. We consider up to five interferometers (H: LIGO-Hanford, L: 
LIGO-Livingston, V: Virgo, K: KAGRA, I: LIGO-India). All of them are assumed to 
have an identical noise curve.  The networks with bold letters are those shown 
in Fig.6.}
  \begin{tabular}{|l|c|c|c|c|c|} \hline
    network & $s_2$ & $s_4$ & $s_6$ & $s_{8}$ & $s_{10}$ \\ \hline 
   {\bf single} & 1 & 1 & 1& 1 &1 \\
   {\bf HL} & 0.910762 & 0.846346 & 0.795697& 0.753916 &0.718351 \\
   HV & 0.560663 & 0.410021 & 0.334791& 0.289303 &0.258356 \\
   VK & 0.721165 & 0.587502 &0.504867 &0.447881 &0.405944 \\
  {\bf   HLV} & 0.587682 & 0.421862 & 0.329701& 0.27037 &0.228792 \\
 {\bf  HLVK} & 0.495651 & 0.311737 & 0.221355& 0.169049 &0.135311 \\
 HLVI & 0.470593 & 0.29952 & 0.217136& 0.168838 &0.137002 \\
 {\bf  HLVKI} & 0.425877 & 0.246289 & 0.164202& 0.118532 &0.089929 \\ \hline
  \end{tabular}
\end{table}

\begin{figure}
\begin{center}
\includegraphics[width=8.cm,clip]{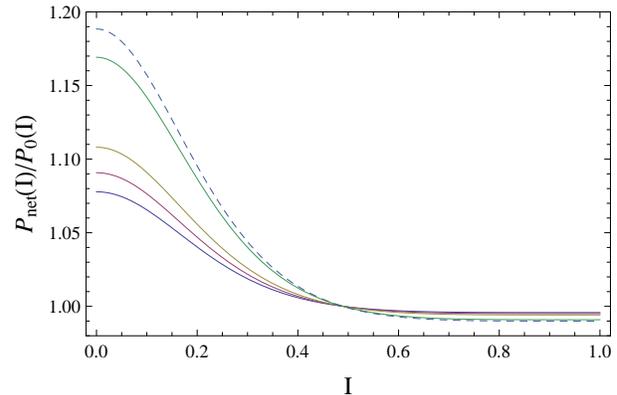}
\end{center}

\vspace*{-0.2cm}

\caption{The ratios $P_{net}(I)/P_0(I)$ for various networks. Dashed 
curve is given  for a single detector network and four solid curves are for 
HL, HLV, HLVK
and HLVKI (from top to bottom at $I=0$). 
}
\label{t3}
\end{figure}
\subsection{Total Detection Rate}

So far, we have studied  PDFs of inclinations $I$ (and $\theta$). In this 
subsection,  we go back to \S2.3 about the relative detection rate which was analytically 
examined in Seto (2014).
We apply  our perturbative formulation for the ratio $Y=X/X_0$ defined in 
Eq.(28).  This ratio represents validity of Schutz's approximation $X_0$ for 
estimating the relative 
detection rates $X$.

From Eqs.(\ref{mnet0}) and (\ref{mnet})
 we can easily obtain
\beq
Y=\frac{M_{net}}{N_0}=1+\frac{3}{16} \frac{u_2 s_2}{N_0} +\frac{9}{1024} \frac{u_4 s_4}{N_0}+\frac{35}{16384} \frac{u_6 s_6}{N_0}+\cdots,
\eeq
and the ratio $Y$ can be directly evaluated, as actual numbers. In Table.4, we 
provide them  for various networks of detectors, again assuming that all the component
detectors have the same sensitivity.

As expected from Table 3, the HL network has the deviation $0.92\%$ close to the 
maximum value $1.01\%$ for a single detector (see also inequalities (39)).  
This deviation would be sufficiently small for astronomical arguments, but the 
deviation for the HLVKI network is further smaller and $\sim 0.42\%$.

\begin{table}
\caption{The ratio $Y$ for various networks}
  \begin{tabular}{|l|c|c|c|c|} \hline
    single & HL & HLV & HLVK & HLVKI  \\ \hline 
 1.010125 & 1.00919 & 1.00588& 1.00495 &1.00424 \\ \hline
  \end{tabular}
\end{table}

\section{Summary}

In this paper, we discussed the probability distribution function $P_{net}(I)$  of inclinations $I=\cos\theta$ ($\theta$: 
inclination angle) for compact 
binaries that are detected by a coherent signal analysis with a network of ground-based 
GW interferometers. In a coherent signal analysis, the SNR of a binary depends 
not only on its sky direction $\ven$ and inclination $I$, but also on 
its polarization angle $\psi$. We have extensively used the simple form 
(\ref{deff}) given by Cutler and Flanagan (1994) to properly include the 
$\psi$-dependence.  Here we have an important parameter    $\eps(\ven)$ that  characterizes the asymmetry of the network 
sensitivities to two orthogonal polarization modes from direction $\ven$. This 
parameter 
has the identity $\eps(\ven)=1$ for a single interferometer and an asymptotic 
behaviour $\eps(\ven)\to 0$ for large number of randomly placed interferometers.
One of the central issues in this paper was how to deal with the effects of the 
parameter $\eps(\ven)$.

Schutz (2011) derived a PDF  
under a simplification equivalent to setting
$\eps(\ven)=0$ in this paper.  This simplified  PDF corresponds to $P_0(I)$ defined in 
Eq.(\ref{p0}), and we showed that it works well 
 for face-on binaries ($I=1$) with errors less than $\sim1\%$. On the other hand, 
 for edge-on binaries ($I=0$), this function is $\sim 20\%$ smaller than 
 $P_1(I)$ defined for a single 
 interferometer.

 In the cumulative form defined in Eq.(\ref{cnet}), the PDF for a given network
 is reproduced by the simple expression ${\cal P}_{cum}(\theta,0)$
 at $\sim 1\%$ accuracy.  Therefore, 
the fitting formula (\ref{pcumf}) and Table.1 would be useful for  astronomical 
arguments such as  
prospects of EM counterpart searches triggered by  GW detections.

We also developed a perturbative method to evaluate the function $P_{net}(I)$ by 
introducing the network parameters $s_j$ ($j=2,4,,\cdots$). These parameters are given by  certain 
angular averages of the moments $\eps(\ven)^j$. Convergence of our expansion is 
fast, and  expressions including the first few correction terms of 
Eqs.(\ref{qnet}) and (\ref{mnet}) would be sufficient in practice. Even if the 
horizon distances of individual interferometers are different, we can easily 
apply our method for arbitrary networks, by introducing appropriate weights for detectors.

We generated the PDFs concretely for the potential networks composed by the 
second generation detectors.
The network with the two LIGO interferometers (HL) has relatively large values 
$s_j$, due to their nearly aligned configurations, and the function $P_{net}(I)$ 
is similar to $P_1(I)$ defined for a single interferometer.  On  the other hand, the PDF of the network 
composed by the five interferometers (HLVKI) is closer to $P_0(I)$ with smaller network
parameters $s_j$.

The author thanks to H.Tagoshi and K.Kyutoku for helpful conversations.
This work was supported by JSPS (24540269) and
MEXT (24103006).




\begin{thebibliography}{DUM}



\bibitem[\protect\citeauthoryear{Abadie et al.}{2010}]{2010Class.Quan.Grav..27q3001A} 
Abadie J., et al., 2010, Class.Quan.Grav, 27, 173001 


\bibitem[\protect\citeauthoryear{Apostolatos et 
al.}{1994}]{1994Phys.Rev.D..49.6274A} Apostolatos T.~A., Cutler C., Sussman 
G.~J., Thorne K.~S., 1994, Phys.Rev.D, 49, 6274 


\bibitem[\protect\citeauthoryear{Arun et al.}{2014}]{2014arXiv1403.6917A} 
Arun K.~G., Tagoshi H., Kant Mishra C., Pai A., 2014, arXiv, 
arXiv:1403.6917 

\bibitem[\protect\citeauthoryear{Berger}{2013}]{2013arXiv1311.2603B} Berger 
E., 2013, arXiv, arXiv:1311.2603 

\bibitem[\protect\citeauthoryear{Berger, Fong, 
\& Chornock}{2013}]{2013ApJ...774L..23B} Berger E., Fong W., Chornock R., 2013, ApJ, 774, L23 


\bibitem[\protect\citeauthoryear{Blanchet et 
al.}{2008}]{2008Class.Quan.Grav..25p5003B} Blanchet L., Faye G., Iyer B.~R., Sinha 
S., 2008, Class.Quan.Grav, 25, 165003 


\bibitem[\protect\citeauthoryear{Cannon et al.}{2012}]{2012ApJ...748..136C} 
Cannon K., et al., 2012, ApJ, 748, 136 



\bibitem[\protect\citeauthoryear{Cutler 
\& Flanagan}{1994}]{1994Phys.Rev.D..49.2658C} Cutler C., Flanagan {\'E}.~E., 1994, Phys.Rev.D, 49, 2658 


\bibitem[\protect\citeauthoryear{Dietz et al.}{2013}]{2013Phys.Rev.D..87f4033D} 
Dietz A., Fotopoulos N., Singer L., Cutler C., 2013, Phys.Rev.D, 87, 064033 


\bibitem[\protect\citeauthoryear{Evans et al.}{2012}]{2012ApJS..203...28E} 
Evans P.~A., et al., 2012, ApJS, 203, 28 



\bibitem[\protect\citeauthoryear{Fairhurst}{2011}]{2011Class.Quan.Grav..28j5021F} 
Fairhurst S., 2011, Class.Quan.Grav, 28, 105021 



\bibitem[\protect\citeauthoryear{Finn 
\& Chernoff}{1993}]{1993Phys.Rev.D..47.2198F} Finn L.~S., Chernoff D.~F., 1993, Phys.Rev.D, 47, 2198 


\bibitem[\protect\citeauthoryear{Ghosh 
\& Bose}{2013}]{2013arXiv1308.6081G} Ghosh S., Bose S., 2013, arXiv, arXiv:1308.6081 



\bibitem[\protect\citeauthoryear{Hotokezaka et 
al.}{2013}]{2013ApJ...778L..16H} Hotokezaka K., Kyutoku K., Tanaka M., 
Kiuchi K., Sekiguchi Y., Shibata M., Wanajo S., 2013, ApJ, 778, L16 





\bibitem[\protect\citeauthoryear{Kelley, Mandel, 
\& Ramirez-Ruiz}{2013}]{2013Phys.Rev.D..87l3004K} Kelley L.~Z., Mandel I., Ramirez-Ruiz E., 2013, Phys.Rev.D, 87, 123004 


\bibitem[\protect\citeauthoryear{Kyutoku, Ioka, 
\& Shibata}{2014}]{2014MNRAS.437L...6K} Kyutoku K., Ioka K., Shibata M., 2014, MNRAS, 437, L6 


\bibitem[\protect\citeauthoryear{Kyutoku 
\& Seto}{2014}]{2014MNRAS.441.1934K} Kyutoku K., Seto N., 2014, MNRAS, 441, 1934 


\bibitem[\protect\citeauthoryear{LIGO Scientific Collaboration et 
al.}{2013}]{2013arXiv1304.0670L} LIGO Scientific Collaboration, et al., 
2013, arXiv, arXiv:1304.067


\bibitem[\protect\citeauthoryear{Metzger 
\& Berger}{2012}]{2012ApJ...746...48M} Metzger B.~D., Berger E., 2012, ApJ, 746, 48 


\bibitem[\protect\citeauthoryear{Nakar}{2007}]{2007PhR...442..166N} Nakar 
E., 2007, Phy.Rep, 442, 166 



\bibitem[\protect\citeauthoryear{Nissanke et 
al.}{2010}]{2010ApJ...725..496N} Nissanke S., Holz D.~E., Hughes S.~A., 
Dalal N., Sievers J.~L., 2010, ApJ, 725, 496 


\bibitem[\protect\citeauthoryear{Nissanke, Kasliwal, 
\& Georgieva}{2013}]{2013ApJ...767..124N} Nissanke S., Kasliwal M., Georgieva A., 2013, ApJ, 767, 124 

\bibitem[\protect\citeauthoryear{Peters 
\& Mathews}{1963}]{1963Phys.Rev...131..435P} Peters P.~C., Mathews J., 1963, Phys.Rev., 131, 435 



\bibitem[\protect\citeauthoryear{Piran, Nakar, 
\& Rosswog}{2013}]{2013MNRAS.430.2121P} Piran T., Nakar E., Rosswog S., 2013, MNRAS, 430, 2121 



\bibitem[\protect\citeauthoryear{Sathyaprakash 
\& Schutz}{2009}]{2009LRR....12....2S} Sathyaprakash B.~S., Schutz B.~F., 2009, Liv.Rev.Rel, 12, 2 



\bibitem[\protect\citeauthoryear{Schutz}{2011}]{2011Class.Quan.Grav..28l5023S} Schutz 
B.~F., 2011, Class.Quan.Grav, 28, 125023 




\bibitem[\protect\citeauthoryear{Seto}{2014}]{seto} Seto, N., 2014, arXiv, arXiv:1406.4238




\bibitem[\protect\citeauthoryear{Tagoshi et 
al.}{2014}]{2014arXiv1403.6915T} Tagoshi H., Kant Mishra C., Pai A., Arun 
K.~G., 2014, arXiv, arXiv:1403.6915 

\bibitem[\protect\citeauthoryear{Tanvir et al.}{2013}]{2013Natur.500..547T} 
Tanvir N.~R., Levan A.~J., Fruchter A.~S., Hjorth J., Hounsell R.~A., 
Wiersema K., Tunnicliffe R.~L., 2013, Natur, 500, 547 



\bibitem[\protect\citeauthoryear{Van Den Broeck 
\& Sengupta}{2007}]{2007Class.Quan.Grav..24..155V} Van Den Broeck C., Sengupta A.~S., 2007, Class.Quan.Grav, 24, 155 




\end{thebibliography}
\end{document}